\documentclass[twocolumn,aps,pra]{revtex4-1}

\usepackage{amssymb} \usepackage{amsfonts} \usepackage{amsmath}
\usepackage{graphicx} \usepackage{bbm}
\usepackage{color}

\begin{document}

\author{Sven Gnutzmann$^1$, Holger Schanz$^{2,3}$ and Uzy
  Smilansky$^{4,5}$}

\affiliation {$^{1}$School of Mathematical Sciences,
  University of Nottingham, Nottingham NG7 2RD, UK \\
  $^{2}$Institute for Mechanical Engineering, Hochschule
  Magdeburg-Stendal,
  39114 Magdeburg, Germany\\
  $^{3}$Max-Planck-Institute for Physics of Complex Systems, D-01187 Dresden, Germany\\
  $^{4}$School of Mathematics, University of Cardiff, Cardiff CF24 4AG, UK\\
  $^{5}$Department of Physics of Complex Systems, The Weizmann
  Institute of Science, Rehovot, Israel}

\title{Topological Resonances in Scattering on Networks (Graphs)}

\begin{abstract}
  We report on a hitherto unnoticed type of resonances occurring in
  scattering from networks (quantum graphs) which are due to the
  complex connectivity of the graph - its topology.
  We consider generic open graphs and show that any cycle leads to narrow resonances which
  do not fit in any of the prominent
  paradigms for narrow resonances
  (classical barriers, localization due to disorder, chaotic
  scattering). We call these resonances `topological' to emphasize their origin
  in the non-trivial connectivity.
  Topological resonances have a clear and unique signature which is apparent in the
  statistics of the resonance parameters
  (such as e.g., the width, the delay time or the
  wave-function intensity in the graph). We discuss this phenomenon by providing analytical arguments
 supported by numerical simulation, and identify
  the  features of the above distributions which
  depend  on genuine topological quantities such as
  the length of the shortest cycle (girth).
  These signatures cannot be explained using any of the other paradigms for
  narrow resonances.
  Finally, we  propose an experimental setting
  where the  topological resonances could be demonstrated, and study the
  stability of the relevant distribution functions to moderate
  dissipation.
\end{abstract}


\maketitle

Narrow resonances are abundant in a large variety of physical systems, and their immense
importance as indicators of long lived states led to an intensive study of their
properties and origin.
The weak effective coupling to the continuum which underlies
their appearance can arise in
various circumstances.
Common mechanisms are the existence of potential or
dynamical barriers such as weakly transmitting optical mirrors in optical
Fabry-Perot resonators, dynamical tunneling in systems with a mixed
phase space \cite{HKL00,M+08} or scarring of quantum states by unstable
trapped orbits \cite{BAF01}.
Anderson localization in disordered systems
\cite{AL1,AL2} induces narrow resonances not because of any barriers but
because of destructive interference between multiply scattered waves with random phases
while a classical particle diffuses unhindered through the system. Here,
 the resonance parameters depend on the particular realization of the
disorder and must be studied with statistical methods.
Statistical methods are also necessary
for studying resonances which characterize chaotic
scattering \cite{LW91,FS97,SFT99,sc1,sc2,novaes}, where fluctuations in the wave functions may lead to
approximate bound states with very low amplitude at the interface between the
interior of the system and the continuum. This inhibits the transition to the
exterior, resulting in a long living state.

Wave propagation on bounded networks (graphs) displays many features
which are typical to quantum chaotic systems
\cite{KS97,GS03}. When the networks are connected to
external leads, the resulting scattering parameters fluctuate, much in the same way as expected
from the analysis of chaotic wave-scattering in open Hamiltonian systems \cite {sc1,sc2,KS00,KS03}.
However, in addition to resonances from random-like interfering waves, the non-trivial connectivity is also
responsible for the formation of another type of narrow resonances --
the subject of the present note.
We will show that these resonances exist in a large class of
scattering graphs if the graph contains a cycle. We thus call them
`topological resonances'. They have properties which
clearly distinguish  them from other mechanisms leading to narrow resonances:
their mark on the  distributions of the resonance parameters
cannot be explained by any of the other paradigms.

The rest of this letter is organized as follows. The topological resonances signature will first be illustrated
with some numerical simulations. The underlying theoretical framework,
will then be reviewed, and used to derive the observed resonance distributions in
simple cases. Finally, a possible experimental setup is proposed
where topological resonances could be observed.

\begin{figure}[htb]
\begin{center}
\includegraphics[width=0.45\textwidth]{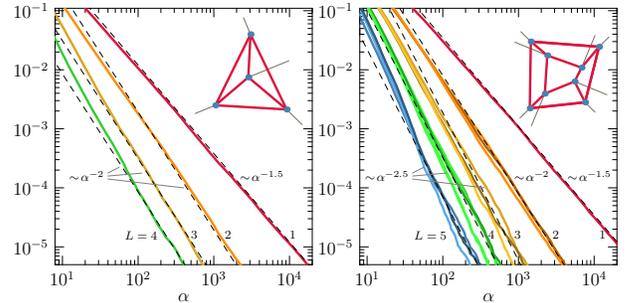}
\end{center}
 \caption{\label{fig1} (Color online)
  Tails of the
   integrated distribution $I(\alpha)$
   for the mean intensity $\alpha$.
   Insets show  the underlying
   graphs.
   Full lines represent numerical results where
   $L=1,\dots,5$
   is the number of attached lead. Note that the data
   for  $L=2,3,4,5$ for the cube consists of more than one line
   -- these correspond to non-equivalent ways
   to attach $L$ leads.
   The dashed lines are power-law tails $\sim \alpha^{-\mu}$
   with indicated exponents.
}
\end{figure}

\emph{Numerical illustrations:}  The insets of Fig.~\ref{fig1} show two different networks: a fully
connected graph with 4 vertices (tetrahedron) on the left and a graph with 8
vertices (cube) on the right. The interior of the graph consists of
finite bonds  between vertex pairs (red lines).
Infinite leads (gray lines) are attached to some
vertices.
Scalar waves propagate
freely on the bonds and leads, and at the vertices they are reflected
or transmitted without losses.
Thus, one has a scattering system, which is described by an $L\times L$ scattering matrix $S(k)$.

Solving the scattering problem numerically with an
incoming wave of unit flux and  wave number $k$, we
compute the mean intensity $\alpha(k)$  of the wave function on the internal
edges (see \eqref{alphak} for an explicit definition). The resulting distributions
\begin{equation}
  P(\alpha)= \lim_{K \to \infty }\frac{1}{K} \int^{K}
  \delta(\alpha - \alpha(k)) dk\  ,
  \label{Palpha}
\end{equation}
or rather, their cumulative form
$I(\alpha)=\int_\alpha^\infty P(\alpha')d \alpha'$, display extremal power-law distributions shown in 
in Fig.~\ref{fig1},  for various $L$
values. 
The simulation summarized in Fig.~\ref{fig1} and further numerical
results for other graphs suggest $ I(\alpha)\sim \alpha^{-\mu}$ for large $\alpha$ with
\begin{equation}\label{exponent}
  \mu=
  \begin{cases}
    \frac{L+2}{2} & \text{for $L\le C-1$}\\
    \frac{C+1}{2} & \text{for $L \ge C-1$}
  \end{cases}
\end{equation}
where $C$ is the  number of bonds
on the shortest cycle (girth).
The tetrahedron girth is $C=3$. Thus, one
expects the exponent to be $\mu_{L}=1.5$ for $L=1$ and $\mu_L=2$ for $L\ge
2$. On the other hand, for the cube with $C=4$ one expects
$\mu_{1}=1.5$, $\mu_{2}=2$, and $\mu_{L}=2.5$ for $L\ge 2$. This is indeed
borne out by the simulations. The appearance of the girth suggests
a topological origin.

The result \eqref{exponent} cannot be explained
by the other paradigms for narrow resonances.
Indeed barriers such as almost perfect mirrors in a Fabry-Perot interferometer
lead to a cut-off for the intensity. Scattering from a
disordered system gives rise to a power
law with a fixed exponent
$\mu_{\mathrm{loc}}=1$ which is independent of the number of channels \cite{AL1,AL2}.
Random-matrix models for chaotic scattering
predict a power law with an exponent $\mu_{\mathrm{RMT}}=\frac{L+2}{2}$
\cite{LW91,FS97,SFT99}
which is only consistent with \eqref{exponent} if $L\le C-1$.

We shall now summarize the graph theoretical setting
(including the definition of the class of graphs)
where topological resonances will be rigorously defined,
and equation
\eqref{exponent} will be derived.

\emph{Quantum Graphs and topological resonances:}
A scattering graph
$G=\{\mathcal{V},\mathcal{E}\}$ consists of a set of vertices $\mathcal{V}$
and a set of edges $\mathcal{E}=\mathcal{B} \cup \mathcal{L}$ where
$\mathcal{B}$ is the set of bonds connecting pairs of vertices and $\mathcal{L}$
the set of infinite leads.
The graphs considered here are all of finite cardinality.
We will assume that the graph is connected, has no loops (each bond
connects two different vertices) and each vertex $i$
has degree
(the number of attached edges)
$d_i \ge 2$. Thus, graphs with dangling bonds corresponding to vertices with $d=1$ are excluded.
Multiple connections between two vertices are allowed.
Each bond $b \in \mathcal{B}$ has a finite
length $\ell_b \in (0,\infty)$ and a coordinate $x_b\in [0,\ell_b]$ such that
$x_b=0$ and $x_b=\ell_b$ correspond to the two end vertices.
The leads $l \in
\mathcal{L}$ are of infinite length. The coordinate $x_l \in [0, \infty)$ is
defined such that $x_l=0$ is the position of the end vertex.

The complex
valued  wave function on the graph is written as
$ \Psi = \left\{
  \psi_b(x_b) \right\}_{b \in \mathcal{B}} \cup \left\{ \psi_l(x_l)
\right\}_{l \in \mathcal{L}} \equiv \left\{\psi_e(x_e)\right\}_{e \in
  \mathcal{E}} $.
On each edge the wave function satisfies the Helmholtz
(or stationary free Schr\"odinger) equation $\frac{d^2 \psi_e}{dx_e^2}+ k^2
\psi_e=0$ where $k>0$ is the wave number. At the vertices the wave
function is continuous and $\sum_{e=1}^d \frac{d \psi_e}{dx_e} (0) = 0$,
where the sum extends over all edges which emanate from the vertex.  These
Neumann or Kirchhoff matching conditions are
a standard choice from a wider range of admissible boundary conditions
\cite{KosSch}.
At a given wave number the wave function on any edge has the
form
\begin{equation}
  \psi_e(x_e)= a_{(e,+)} e^{ik x_e} + a_{(e,-)} e^{-ikx_e}
  \label{plane_waves}
\end{equation}
where $a_{(e,-)}$  and $a_{(e,+)}$ are the amplitudes
of the two counter-propagating waves on the edge.
The mean intensity $\alpha(k)$ on the graph is defined by
\begin{equation}
  \alpha(k)=\frac{1}{\left| \mathcal{B}\right|}\sum_{b \in \mathcal{B};\sigma=\pm}
  |a_{b,\sigma}|^2\ .
  \label{alphak}
\end{equation}

If the bond lengths are rationally independent then the
spectrum of the quantum graph is purely continuous with
generalised eigenstates which are bounded everywhere
but are not normalizable. To each value of $k>0$ one can associate a unitary
scattering matrix $S(k)$ \cite{KS00} that relates the outgoing
coefficients
on the leads to the incoming ones.  Resonances are identified as poles
of the scattering matrix when $k$ is in the upper complex $k$-plane.
The (positive) imaginary part of the wave number at a resonance
gives the decay rate (width).
If the bond lengths are
changed continuously so do the positions of the poles of $S(k)$.
When bond lengths become rationally dependent  some poles may move to the real axis
indicating the appearance of a normalizable bound state embedded in
the continuum
(see \cite{kuchment,PhD,Davies}).

We can now \emph{define} a topological resonance as a pole of the
scattering
matrix that
 can be moved to the real line to form a
bound state by changing some bond lengths continuously while the graph connectivity and
matching conditions remain unchanged.

The main statement of this letter is that
for rationally independent bond lengths a quantum graph as described above
supports topological resonances if and only if
it contains a cycle.

\begin{figure}[htb]
\begin{center}
\includegraphics[width=0.45\textwidth]{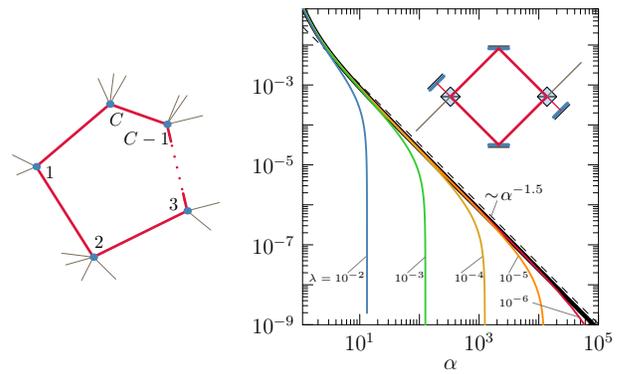}
\end{center}
 \caption{\label{fig2} (Color online)
   Left: A cycle of length $C$.\\
   Right: Numerically obtained tail of the integrated distribution $I(\alpha)$
   of the mean intensity for a lossy
   beam-splitter setup (see inset).
}
\end{figure}

This can be shown following similar ideas
as in
\cite{SK03}. Let us first consider any scattering graph that does not
contain a cycle. Such a graph is a tree (and to satisfy the requirement $d\ge 2$ all the canopy edges must be leads).
The wave function for a bound state has to vanish on all the leads
in order to be square-integrable. 
The remaining edges (bonds) form a finite tree whose leaves are at one
end only connected to leads.
The matching conditions then imply
that the wave function also has to vanish on all the leaves. The process can now be iterated on the remaining tree,
showing that there are no non-vanishing square-integrable
solutions on a scattering tree graph (of finite cardinality).
As a result there cannot be any topological resonances on a scattering
graph without cycles.\\
Now let us assume that the scattering graph contains a cycle and let
us show that there are choices for the bond lengths that lead to bound
states which we  call \emph{topological bound states}. Topological
resonances are the remnants of topological bound states when the
latter get mixed with the continuum by a generic choice of bond
lengths.  The mechanism is related to a similar phenomenon
for closed graphs
where it explains the structure of \emph{scars} \cite{SK03}.
 Let $\mathcal{C}$ be a cycle in the
graph that consists of $C=|\mathcal{C}|$ bonds as shown on the left in
Fig.~\ref{fig2}. Our assumptions (no loops) imply $C\ge2$.
 Now let us construct a bound state on
$\mathcal{C}$. We require that the wave function vanishes exactly
outside $\mathcal{C}$. By continuity all vertices on $\mathcal{C}$ are
then nodal points.  The wave function for any $ b \in \mathcal{C}$
then has to be of the form $\pm \sin (k x_b)$ and the matching
conditions reduce to the statement that the wave function and its
derivative have to be continuous along the cycle. This implies the
following two diophantine conditions:
\begin{align}
  k \ell_b=& n_b \times \pi&& \text{with $n_b \in \mathbb{N}$ for all
    $b \in \mathcal{C}$}
  \label{topological_bound_state_1}
  \\
  \sum_{b \in \mathcal{C}} n_b = & 2 s && \text{for some $s \in
    \mathbb{N}$.}
  \label{topological_bound_state_2}
\end{align}
The above conditions can be satisfied for a discrete sequence of wave
 numbers if and
only if the bond lengths on $\mathcal{C}$ are rationally
dependent. This means that there exists a unit of length $\ell_0>0$
such that
$
  \ell_b = i_b \ell_0 $  for some $i_b\in \mathbb{N}$.
One finds topological bound states on the cycle for wave numbers $n
k_0$ ($2nk_0$) with $n \in \mathbb{N}$ if $\sum_{b\in \mathcal{C}}
i_b$ is even (odd).
The only reason for the existence of such states
is the combination of a
topological structure (the cycle) with destructive interference at the
nodes.\\
A generic (rationally independent) choice of bond lengths will destroy
topological bound states on the cycle as the conditions
\eqref{topological_bound_state_1} and
\eqref{topological_bound_state_2} cannot be satisfied exactly.
However the condition can be satisfied approximately -- to arbitrary
precision at appropriate wave numbers.  This is where the destroyed
topological bound
states leave a mark in the form of topological resonances.

Let us now show that topological resonances on cycles
indeed lead to the statistical signatures that we observed
in the numerical simulations reported above.
For this we consider a connected scattering graph with girth $C\ge 2$
and focus our attention on the corresponding cycle. Each vertex has two attached
bonds which belong to the cycle and at least one additional attached
edge that does not belong to the cycle (as shown in the left part of
Fig.~\ref{fig2}).
We will show that the
ratio $\rho$ of the intensity inside the cycle to the intensity outside
has a distribution with a power-law tail $P(\rho)\sim
 {\rho}^{-\mu_\mathcal{C}-1}$
with $\mu_{\mathcal{C}}=\frac{C+1}{2}$.
For simplicity we will assume that all vertices on $\mathcal{C}$ have
the same degree $d=3$ (the calculation for the general case follows
the same steps but is too cumbersome for this note).
Since we only want to compare the intensity inside the
cycle to the intensity on the edges that are adjacent to the vertices
on the cycle we may disregard the rest of the graph and replace the
adjacent edges by leads of infinite length. We have thus reduced the
problem to finding the scattering solutions for one cycle of length
$C$ with one lead at each vertex.
The mean intensity on the cycle is defined as in \eqref{alphak} where
the sum is restricted to bonds on $\mathcal{C}$.
Combining the amplitudes
in one vector $\mathbf{a}_{\mathcal{C}}$ one finds (see
\cite{KS00,KS03})
\begin{equation}
  \mathbf{a}_{\mathcal{C}} = \left(\mathbbm{1}-
    e^{ik \ell/2} \sigma_{\mathcal{CC}} e^{ik \ell/2}\right)^{-1} e^{ik \ell/2}
  \sigma_{\mathcal{CL}}  \mathbf{b}_{\mathcal{L}}^{\mathrm{in}}
  \label{amplitude}
\end{equation}
Here $\ell$ is a diagonal matrix of size $2C\times 2C$ that contains
each bond length of the cycle twice,
 $\mathbf{b}_{\mathcal{L}}^{\mathrm{in}}$ are amplitudes of
incoming waves on the leads such that the mean intensity outside the
cycle is proportional to $|\mathbf{b}_{\mathcal{L}}^{\mathrm{in}}|^2$
(flux conservation ensures that the outgoing waves have the same
intensity).
The $2C \times 2C$ matrix $\sigma_{\mathcal{CC}}$ and the $2C \times
C$ matrix $\sigma_{\mathcal{CL}}$ contain scattering amplitudes that can be
derived from the Neumann matching conditions at the vertices.
They are given by
\begin{equation}
  \sigma_{\mathcal{CL}}= \frac{2}{3}
  \begin{pmatrix}
    P\\
    \mathbbm{1}
  \end{pmatrix} ;\qquad
  \sigma_{\mathcal{CC}}= \frac{1}{3}
  \begin{pmatrix}
    2 P & - \mathbbm{1}\\
    -\mathbbm{1} & 2 P^T
  \end{pmatrix}
\end{equation}
where $P$ is the permutation matrix for the cyclic permutation
$(12\dots C)$. The sub-unitary matrix $\sigma_{\mathcal{CC}}$ has one
eigenvalue equal to one with the eigenvector $\mathbf{v}_0=\left(
  \begin{smallmatrix}
    \mathbf{1}_C^T \\
    -\mathbf{1}_C^T
  \end{smallmatrix}\right)^T
$ where $\mathbf{1}_C$ is the $C$-dimensional vector with unit
entries.
Bound states appear whenever
\begin{equation}
\Sigma(k)=e^{ik\ell/2}
\sigma_{\mathcal{CC}} e^{ik\ell/2}
\end{equation}
has an eigenvalue equal to
one.
This happens if $e^{ik\ell/2}=\mathbbm{1}$ which
cannot be satisfied for generic (rationally independent)
bond lengths and $k>0$.
However, in that case $k \mapsto e^{i k \ell/2}$ is an ergodic
flow on a $C$-torus and $e^{ik\ell/2}=\mathbbm{1}$ defines a point on
the torus that can be approached to arbitrary precision as $k$
increases\cite{Barra}.  Defining the ratio of intensities as
$\rho(k)= |\mathbf{a}_{\mathcal{C}}|^2/|\mathbf{b}_{\mathcal{L}}^{\mathrm{in}}|^2$
we may derive a power law tail for $P(\rho)=\langle
\delta(\rho-\rho(k))\rangle_k$
by
replacing the spectral average by a torus average
$P(\rho)=(2\pi)^{-C}\int d^{C}\boldsymbol{\theta} \delta (\alpha
-\alpha(\boldsymbol{\theta}))$
where we have replaced the $2C\times
2C$-matrix $e^{i k\ell}$ by a diagonal matrix
$e^{i\boldsymbol{\theta}}$ that parameterizes the $C$-torus (each
angle $\theta_j$ appears twice).  Focussing on the contribution
from a small region around $\boldsymbol{\theta}=\mathbf{0}$ where
$\Sigma(\boldsymbol{\theta})$ has an eigenvalue one which dominates
the behaviour. Second-order perturbation then yields
\begin{equation}
  \rho(\boldsymbol{\theta})\sim \frac{f^{(2)}(\boldsymbol{\hat{\theta}})}{\bar{\theta}^2+
    [g^{(2)}(\boldsymbol{\hat{\theta}})]^2}
\end{equation}
where $\overline{\theta}= \frac{1}{C} \sum_{n=1}^C \theta_n$ and
$\hat{\theta}_n= \theta_n-\overline{\theta}$.  The functions
$f^{(2)}(\boldsymbol{\hat{\theta}})$ and
$g^{(2)}(\boldsymbol{\hat{\theta}})$ are (explicitly known \cite{GSS13}) positive
definite quadratic forms in the variables $\hat{\theta}_n$.  This
implies $P(\rho)\sim \rho^{-\mu_{\mathcal{C}}-1}$ with $\mu_{\mathcal{C}}=\frac{C+1}{2}$.\\
Coming back to the full graph that contains $\mathcal{C}$ as a
subgraph, note that the mean intensity $\alpha$
on the graph contains a contribution proportional to $\rho$ from
the cycle. As a consequence the tail of $P(\alpha)$ cannot decay
faster than the tail of $P(\rho)$ such that
$\mu_{\mathcal{C}}$ gives a lower bound for the exponent $\mu$. This is
consistent
with our numerical findings \eqref{exponent} if $L\ge
C-1$. The exponent $\mu=\frac{L+2}{2}$ for $L< C-1$ is consistent with
the random-matrix approach and can be derived as a lower bound for the
exponent in the present context following similar ideas
\cite{LW91,FS97,SFT99}. Note however that our simulations were performed
on  regular structures where the connectivity does not vary
strongly
and where at most one lead was connected to one vertex.
If we want \eqref{exponent} to be true in other circumstances  we need to redefine $L$
appropriately. E.g. one may show that if there are many leads at the
same vertex with Neumann matching conditions
there is still only one open  quantum channel that
couples to the inside of the graph. Moreover if a graph has a
(possibly large) subgraph that is weakly connected to the rest
(e.g. via a single bridge) then there is only a small number of channels
which connect to the subgraph. The following definition will take care
of these issues. Consider a connected subgraph $H$. A vertex $v$ is on the
boundary $\partial H$ of that subgraph if it is adjacent to at least
one edge in $H$ and at least one edge outside $H$. We redefine
$L$ as the minimum of the size (cardinality) of $\partial H$
over all connected subgraphs $H$ that contain a cycle
and that contain no lead. This reduces to the number of attached leads
for the numerical simulations presented before. With this
definition we conjecture that the result \eqref{exponent}
we have found remains true for generic scattering graphs
with Neumann matching conditions.
We have excluded loops as they always lead to topological bound states
that cannot be destroyed by changing the length of the loop. One may
allow loops if one defines $C$ as the shortest cycle which is not a loop.
Dangling bonds have been excluded because they lead
to a further set of topological resonances on paths between two
vertices with degree one (i.e. between two vertices with mirror-like
reflection) --
\eqref{exponent} will remain correct
if $C$ is redefined appropriately  \cite{GSS13}.
Let us also mention that measuring the intensity on one point of the
graph
will in general give a different but predictable power-law exponent --
they
are related to the local topological structure (e.g. the smallest cycle that
contains the point) rather than the global topology.
Only very regular graphs like the ones we used
in Fig.~\ref{fig1} have the same power-law exponent at every point in
the graph (and thus also for the mean intensity). However, other global quantities which are used to
characterize resonances, such as e.g., the Wigner delay time or the resonance widths can be shown 
to distribute as the parameter $\alpha$ which was used here to render the theoretical discussion more transparent. 

The derivation of the exponents \eqref{exponent}
made explicit use of some properties of the Neumann matching
conditions. The derivation can be generalised to other continuous
matching conditions. For non-continuous matching conditions
topological resonances as defined above can be constructed but
more complex topological features are reflected in the exponents
of the corresponding power laws.
In Fig.~\ref{fig2} (right panel) we give numerical evidence
of a topological resonance in a graph  structure with matching conditions
that can be realized  in a laser experiment using beam-splitters and
mirrors.
The numerics show a clear
power-law distribution with exponent $\mu=3/2$ (dashed line). We have also included
loss (e.g. at the reflection from mirrors) -- we characterize the loss
by the parameter $\lambda$ which gives the fraction of photons which
are lost when travelling once through the whole system.
Lossy setups follow the power-law behaviour up to a cut-off
that increases as losses decrease.
Reducing
losses to $\lambda= 10^{-5}$ may be within reach \cite{Micha_Nir}
showing that topological resonances may be observed in experiment.

The narrow resonances in networks play a very significant
role
in the presence of a nonlinearity as present in nonlinear optical wave
guides or active optical fibres.
The enhanced intensity at a topological resonance
amplifies the nonlinearity to such an extent
that the perturbative treatment
breaks down and typical nonlinear effect such hysteresis
appear \cite{GSD11}.   A detailed discussion and classification
of topological resonances is now in preparation \cite{GSS13}.

\begin{acknowledgments}
  SG and HS thank the Weizmann Institute of Science
  for hospitality.
  This work has been supported the EPSRC research  network `Analysis
  on Graphs'  (EP/I038217/1).
  We would like to thank Nir Davidson, Micha Nixon, Daniel Waltner, and Jens Bolte for
  fruitful discussion.
\end{acknowledgments}

\end{document}